\documentclass[aps,prb,preprint,showpacs]{revtex4}
\usepackage{epsfig}
\usepackage{graphicx}
\begin{document}

\title{Low-temperature internal friction and thermal conductivity in
plastically deformed metals due to dislocation dipoles and random
stresses}

\author{D.V. Churochkin$^{1}$, S. Sahling$^2$, and V.A. Osipov$^1$}
\email{churoch@thsun1.jinr.ru}

\affiliation{$^1$ Bogoliubov Laboratory of Theoretical Physics,
Joint Institute for Nuclear Research, 141980 Dubna, Moscow region,
Russia\\
$^2$ Institut fur Festkorperphysik, IFP, Technische Universitat
Dresden, D-01069 Dresden, Germany}

\date{\today}
\begin{abstract}
The contribution to the low frequency internal friction and the
thermal conductivity due to optically vibrating edge dislocation
dipoles is calculated within the modified Granato-L\"ucke string
model. The results are compared with the recent experiments on
plastically deformed samples of Al, Ta and Nb at low temperatures.
It is shown that the presence of a reasonable density of optically
vibrating dislocation dipoles provides a good fit to the thermal
conductivity in superconducting samples. At the same time, the
internal friction experiments can not be described within the
standard fluttering string mechanism. We found that the problem
can be solved by assuming random forces acting on the dislocation
dipoles. This gives an additional contribution to the internal
friction which describes well the experimental data at low
temperatures while their contribution to the thermal conductivity
is found to be negligible.

\end{abstract}

\vskip 0.2cm 

\vskip 0.5cm \maketitle

\section{Introduction}

Recently,\cite{Pohl:1999,Sahl:2001,Sahl2:2001,Sahl3:2002,Was:2002}
the low-temperature internal friction, thermal conductivity,
specific heat, and heat release of plastically deformed,
high-purity superconducting crystalline samples of Al, Nb, and Ta
have been experimentally studied and compared with measurements on
amorphous SiO$_{2}$
specimen.\cite{Pohl:1999,Sahl:2001,Sahl2:2001,Sahl3:2002} In
particular, it was established that plastic deformation has a
pronounced effect on the internal friction and the thermal
conductivity. Namely, the value of the internal friction can be
increased by two orders of magnitude over that observed on
annealed samples and it becomes comparable to that of amorphous
SiO$_{2}$. Likewise, the thermal conductivity was found to have
the similar value as that of amorphous
SiO$_{2}$.\cite{Sahl:2001,Sahl2:2001} At the same time, neither
long-time heat release nor additional contribution to the heat
capacity expected for amorphous systems was observed. This finding
indicates that the phonon scattering by dislocations appearing in
crystals under plastic deformations could be of importance.

As is well known, there are two principal mechanisms of phonon
scattering by dislocations: the static strain-field scattering and
the reradiation scattering.\cite{Esh:1949,Nab:1951} The first one
is due to the anharmonicity of the dislocation strain field while
the second one results from a possibility of the sound wave to
induce a dislocation vibration. In this case, the incident energy
will be dissipated as long as the dislocation radiates elastic
waves. This is so-called {\it fluttering} mechanism which is
dynamical in its origin. It is a difficult problem to
experimentally clarify what kind of scattering actually dominates
and especially to obtain a quantitative agreement between the
observed effect and theoretical calculations based on either
static or dynamic models. Moreover, it was shown that both
mechanisms of phonon scattering by an array of single dislocations
are failed in the description of various
experiments.\cite{Sahl:2001,Sahl2:2001,Sahl3:2002,Anderson:1972,Anderson:1979}
For example, the estimated contribution to the thermal
conductivity due to the resonant interaction was found to have a
reasonable agreement with the experimental data only at very short
lengths of dislocation lines (otherwise the resonant frequency
becomes too low). Notice that a similar problem has already
emerged in experiments with plastically deformed
LiF.\cite{Anderson:1972,Anderson:1979} Namely, the measurements of
the thermal conductivity and the ballistic phonon propagation in
deformed LiF at low temperatures show that the obtained phonon
scattering is too strong to be explained by static mechanisms of
phonon-dislocation interaction, but is in rough agreement with
calculations based on a resonant or dynamic interaction with
dislocations which can flutter in the stress field of passing
phonons. The crucial role in explanation of the experimental data
plays an assumption given in Ref.\onlinecite{Kneezel:1982} that a
reasonable density of optically vibrating dislocation dipoles is
present in LiF. In this case, the resonant frequency becomes
markedly higher. It should be mentioned that dislocation dipoles
with a rather high density were actually observed in LiF
\cite{Gilman:1960} as well as in some other materials. For
example, estimates from experiments of the ratio of the
dislocation dipole density to a density of single dislocations
range from $1000$ (in strain-hardening measurements
\cite{Gilman:1960}) to $100$ (by deformation-induced bulk-density
changes \cite{Davidge:1964}) and less (by electron microscopy
\cite{Washburn:1966,Hesse:1972}).

The purpose of this paper is to show that the concept of dynamical
scattering of phonons by dislocation dipoles \cite{Kneezel:1982}
previously used for the explanation of the experiments in LiF
\cite{Anderson:1972,Anderson:1979} could also be suitable for Al,
Nb, and Ta. As was observed in
Refs.\onlinecite{Sahl:2001,Sahl2:2001,Sahl3:2002}, these metals
show a complex dislocation structure under plastic deformation. We
assume that, by analogy with LiF, dislocation dipoles are also
present in plastically deformed Al, Nb, and Ta. It should be noted
that the dipole is a stable dislocation structure arising under
plastic deformation in metals (see, e.g.,
Ref.\onlinecite{Friedel:1964}).
Recently,\cite{Kassner:1999,Kassner:2000} the substructure of the
dislocation ensemble in Al and Cu was experimentally studied under
cyclic deformation. In particular, it was found in
Ref.\onlinecite{Kassner:1999} that for Al deformed at 77 K the
microstructure is composed exclusively of vein dipole bundles and
channels without persistent slip bands. The dislocation density in
the veins is about $3.2\times10^{14}$ m$^{-2}$, and the density in
the channels is about $2.4\times 10^{13}$ m$^{-2}$ with the dipole
separation in a range from 3 nm to 30 nm. It is important to note
that all of the channel defects were found to be edge dipoles. In
addition, the computer modelling performed in
Ref.\onlinecite{Aslanides:2000} shows that there is a critical
size of the dipole separation (1.6 nm for Al and 0.42 nm for Cu)
above which edge-dislocation dipoles in Al and Cu become stable
with respect to the athermal annihilation. Notice that we consider
in our paper dislocation dipoles in Al with a separation close to
20 nm, that is much larger the critical size found in
Ref.\onlinecite{Aslanides:2000}.

Thus, as for LiF the gain in resonant frequency can be achieved
due to optically vibrating dipoles having a reasonable dipole
separation and dislocation length. This allows us to obtain a
quantitative agreement with the experimental data for all three
metals. Evidently, the presence of the fluttering dislocation
dipoles will affect not only the thermal conductivity but also
other physical characteristics, in particular, the internal
friction. We study the contribution to the internal friction due
to vibrating dislocation dipoles in the framework of the modified
Granato-L\"ucke string theory. As is known, the Granato-L\"ucke
vibrating string model is based on an analogy between the
vibration of a pinned dislocation line segment and the forced
damped vibration of a string.\cite{Granato:1956}

Our study shows that a good agreement with the internal friction
experiments in Al, Nb, and Ta \cite{Pohl:1999,Sahl3:2002,Was:2002}
can be obtained only in the presence of some random component of
the stresses. This idea has been recently proposed and studied in
Refs.\onlinecite{Chernov:2002} and \onlinecite{Kamaeva:2004}. In
particular, it was found that the action of random forces on the
dislocation gives a substantial contribution to the decrement. As
was observed experimentally,\cite{Pohl:1999,Sahl3:2002} the
decrement has an unexpectedly large value and practically does not
depend on the amplitude and the frequency of external stresses.
This means that the noise parameters should be governed by the
characteristics of an external signal because this is the only way
to provide the constant contribution to the decrement. Therefore,
we suggest that the external force introduces also a specific time
scale for correlations of stress fluctuations in dislocation
ensemble which always has its own stochastic
dynamics.\cite{Groma:1998} This time scale determines the time of
an additional relaxation of internal stresses due to random
stresses. It seems reasonable to assume that this time is related
to the frequency of the external force. As the simplest
approximation, we use the correlation function exponentially
decreasing with time. What is important, the introduction of a
time scale means that dislocations experience the influence of the
{\it colored} noise. Finally, we adapt the theory proposed in
Refs.\onlinecite{Chernov:2002} and \onlinecite{Kamaeva:2004} to
the case of dislocation dipoles and introduce the additional noise
component with the above-mentioned correlation function in the
equations of motion.

\section{The model}

Let us consider a dislocation dipole with dislocation lines
directed along the z-axis and lying in parallel glide planes at
distance $l$. The Burgers vectors are opposite while the
orientation of dislocation lines is the same. The $xz$ plane is
chosen to be a slip plane for a positive dislocation. In this case
the plane $y=l$ is the slip plane for a negative dislocation.
Within the string approximation, the dislocation dipole can be
modeled by means of two damped interacting vibrating strings.
Supposing that all the parameters of the string are the same for
both dislocations, one can formulate the general equations of
damped glide motion for the edge dislocation dipole in the form
\begin{eqnarray}\label{eq1}
m\frac{\partial ^{2}\epsilon(z,t) }{\partial
t^{2}}+B\frac{\partial\epsilon(z,t)} {\partial
t}-T_0\frac{\partial^{2}\epsilon(z,t)}{\partial z^{2}}=
F^{ext+},\nonumber\\
m\frac{\partial ^{2}\psi(z,t) }{\partial
t^{2}}+B\frac{\partial\psi(z,t)}{\partial t}-
T_0\frac{\partial^{2}\psi(z,t)}{\partial z^{2}}=F^{ext-},
\end{eqnarray}
where $\epsilon(z,t)$ and $\psi(z,t)$ are the displacements of the
positive and negative dislocations, $m$ the effective mass, $T_0$
the line tension, $B$ the damping parameter, and $F^{ext\pm}$ the
total external force which acts on the positive and negative
dislocations, correspondingly, in their glide planes. The
parameters of the model are written as \cite{Kneezel:1982}
\begin{equation}\label{eq3}
m=\frac{\rho b^{2}}{4\pi}[1+(c_{t}/c_{l})^{4}]\ln g, \qquad
T_0=\frac{\mu b^{2}}{4\pi(1-\nu)}\ln g, \qquad B=\frac{\rho
b^{2}\overline{\omega}}{8}[1+(c_{t}/c_{l})^{4}],
\end{equation}
where $\rho$ is the density, $\mu$ the shear modulus,
$\overline{\omega}$ the thermal phonon frequency, $\nu$ the
Poisson constant, $b$ the length of Burgers vector, $c_{t}$ and
$c_{l}$ the transverse and longitudinal sound velocities. At low
frequencies, $g$ can be estimated as $g\sim 1/b\sqrt{\Lambda}$
where $\Lambda$ is the dislocation dipole density. The total
external force, resolved in the glide plane, includes three terms
\begin{eqnarray}\label{eq5}
F^{ext\pm}=f^{\pm}+F^{\pm}+b^{\pm}\eta(t).
\end{eqnarray}
Here $f^{\pm}$ are the interaction forces between the dislocations
in the dipole, $F^{\pm}$ are the forces due to external stress
field $\sigma_{ik}$, and $\eta(t)$ describes a stationary random
component. The explicit form of $\eta(t)$ is not specified at this
stage. Hereafter, the sign plus (minus) corresponds to a positive
(negative) dislocation, and the summation over repeated indices is
assumed. In general, the interaction between dislocations is
determined by the Peach-Koehler force
\begin{equation}\label{eq7}
f^{\pm}_{j}=\varepsilon _{jak}\tau _{a}b_{i}^{\pm} \sigma^{\mp}
_{ik},
\end{equation}
where $\varepsilon _{jak}$ is the totally antisymmetric tensor,
$\sigma^{\mp} _{ik}$ are the corresponding stresses of
dislocations in the dipole, and $\vec{\tau}$ is the unit tangent
vector to the defect line. The displacement fields can be written
as
\begin{equation}\label{eq9}
u_{n}\left(\vec{r},t \right)=\int c_{ijkl}G_{jn,i}\delta
e^{pl}_{kl}dV^{'},
\end{equation}
with $c_{ijkl}$ being the elastic modulus, $G_{jn}$ the Green's
tensor, $G_{jn,i}=\partial G_{jn}/\partial x_i$, and $\delta
e^{pl}_{kl}$ the variation of the plastic part of the strain
tensor. For sliding dislocation one has \cite{Landau:1970}
\begin{eqnarray}\label{eq11}
\delta e^{pl}_{kl}=\frac{1}{2}(b_{k}[\vec{\delta
x}\vec{\tau}]_{l}+ b_{l}[\vec{\delta
x}\vec{\tau}]_{k})\delta(\vec\xi),
\end{eqnarray}
where $\vec{\delta x}$ describes the displacement of the
dislocation line from a straight configuration and
$\delta(\vec\xi)$ is the two-dimensional delta-function. In our
geometry, $\vec b^{+}=-\vec b^{-}=(b,0,0)$, $\vec{\tau}=
(0,0,-1)$, $\vec{\xi}^{+}=(x^{\prime},y^{\prime},0)$, and
$\vec{\xi}^{-}=(x^{\prime},y^{\prime}-l,0)$. The displacements are
chosen to be $\vec{\delta x^{+}}= (\epsilon(z),0,0)$ and
$\vec{\delta x^{-}}=(\psi(z),0,0)$, so that $[\vec{\delta
x^{+}}\vec{\tau}]=(0,\epsilon,0)$ and $[{\vec\delta
x^{-}}\vec{\tau}]=(0,\psi,0)$. After substitution into
Eq.(\ref{eq9}) one obtains
\begin{eqnarray}\label{eq13}
u^{+}_{n}(\vec{r})=\int
c_{ijk2}G_{jn,i}(x,y,z-z^{\prime})b^{+}_{k}\epsilon
(z^{\prime})dz^{\prime},\nonumber\\
u^{-}_{n}(\vec{r})=\int
c_{ijk2}G_{jn,i}(x,y-l,z-z^{\prime})b^{-}_{k}\psi
(z^{\prime})dz^{\prime}.
\end{eqnarray}
For isotropic case,
$$G_{km}(\vec r)=\frac{1}{8\pi\mu}\left[\frac{2}{r}\delta_{km}-
\frac{1}{2(1-\nu)}r_{,km}\right]$$ and $c_{ijkl}=\lambda\delta
_{ij}\delta _{kl}+\mu(\delta _{ik}\delta _{jl}+\delta _{il}\delta
_{jk})$ with $\lambda$ and $\mu$ being the Lame constants.
Supposing that $\epsilon(z)=\epsilon(k)\exp(ikz)$ and
$\psi(z)=\psi(k)\exp(ikz)$, after straightforward calculations one
obtains from Eq.(\ref{eq7})
\begin{equation}\label{eq15}
f^{-}(\vec r)=-2\epsilon(k)e^{ikz}\mu^{2}b^{2}k^{2}
\Biggl[2A\left(1-\frac{8x^{2}y^{2}}{a^{4}}\right)K_{2}(ka)-\frac{4Akx^{2}y^{2}}{a^{3}}K_{1}(ka)+
(2A+B)K_{0}(ka)\Biggr],
\end{equation}
\begin{eqnarray}\label{eq17}
f^{+}(\vec r) &=& -2\psi(k)e^{ikz}\mu^{2}b^{2}k^{2}
\Biggl[2A\left(1-\frac{8x^{2}(y-l)^{2}}{c^{4}}\right)K_{2}(kc)-\frac{4Akx^{2}(y-l)^{2}}{c^{3}}K_{1}(kc)+
\nonumber\\
&&+ (2A+B)K_{0}(kc)\Biggr],
\end{eqnarray}
where $A=-B/4(1-\nu)$, $B=1/4\pi\mu$, $a^{2}=x^{2}+y^{2}$,
$c^{2}=x^{2}+(y-l)^{2}$, and $K_{\nu}$ are the McDonalds
functions. Notice that we are interested only in a component
$f_{1}$ since, at chosen geometry, all the remaining components
become equal to zero in slip planes of dislocations. In the long
wavelength approximation ($kl \ll 1$) one gets
\begin{equation}\label{eq19}
f^{-}(0,l,z)=D\epsilon(z),\quad f^{+}(0,0,z)=D\psi(z),
\end{equation}
where
\begin{equation}\label{eq21}
D=\frac{\mu b^{2}}{2\pi(1-\nu)l^{2}}.
\end{equation}

The second term in Eq.(\ref{eq5}) is also determined by the
Peach-Koehler force which is written in the known form
$F^{\pm}_{r}=\varepsilon _{rak}\tau _{a}b_{i}^{\pm}\sigma _{ik}$
with $\sigma_{ik}$ being an external stress field. In our
geometry,
\begin{equation}\label{eq33}
F^{+}=b^{+}\sigma(0,0,z),\quad F^{-}=b^{-}\sigma(0,l,z).
\end{equation}
Let us consider first the case of free vibrations of the dipole
when $B=F^{\pm}=\eta(t)=0$ in Eq.(\ref{eq1}). There exist both
acoustical and optical modes with $\epsilon=\psi$ and
$\epsilon=-\psi$, respectively. Substituting Eq.(\ref{eq19}) into
Eq.(\ref{eq1}) one obtains
\begin{eqnarray}\label{eq23}
m\frac{\partial ^{2}\epsilon(z,t) }{\partial t^{2}}-
T_0\frac{\partial^{2}\epsilon(z,t)}{\partial z^{2}}\pm
D\epsilon(z)=0.
\end{eqnarray}
Here the sign plus corresponds to the optical mode while the minus
to the acoustical one. The spectrum of normal oscillations is
found to be
\begin{eqnarray}\label{eq25}
\Omega^{\pm}_{n}=\sqrt{\omega^{2}_{n}\pm \frac{D}{m}},
\end{eqnarray}
where $\omega_{n}=(\pi n/L)\sqrt{T_0/m}$ is the resonance
frequency of a single dislocation and L is the dislocation length.
For optical mode, this spectrum was presented in Ref.
\onlinecite{Kneezel:1982}. As is seen, the resonance frequency
becomes higher for the optical mode and the difference increases
with decreasing dipole separation. On the contrary, for acoustical
mode the resonance frequency lies even lower than that for a
single dislocation.

The measured thermal conductivity can be fitted by assuming
dislocation dipoles having a higher resonant frequency and being
more numerous than isolated dislocations.~\cite{Kneezel:1982} For
this reason, we consider below only the case of optically
vibrating dislocation dipoles. Supposing $\epsilon=-\psi$,
$\sigma(0,0,z)= \sigma(0,l,z)$, and taking into account
Eqs.(\ref{eq5}), (\ref{eq19}) and (\ref{eq33}) one can reduce
Eq.(\ref{eq1}) to
\begin{eqnarray}\label{eq35}
m\frac{\partial ^{2}\epsilon(z,t) }{\partial t^{2}}=
T_0\frac{\partial^{2}\epsilon(z,t)}{\partial z^{2}}-
B\frac{\partial\epsilon(z,t)}{\partial
t}+b\sigma(0,0,z)-D\epsilon(z,t)+b\eta(t).
\end{eqnarray}
Let us consider a periodic external stress wave in the form
\begin{eqnarray}\label{eq37}
\sigma(0,0,z)=\sigma_{0}e^{i \omega
t}=\sum\limits_{n}\sigma_{n}\sin (k_{n}z)e^{i \omega t},
\end{eqnarray}
where $\sigma_{0}$ is the shear stress component resolved in the
glide plane, $k_{n}=\pi n/L$, and $\sigma_{n}=4\sigma_{0}/\pi n$
is the Fourier coefficient. Then the general solution to
Eq.(\ref{eq35}) is written as
\begin{eqnarray}\label{ch5}
\epsilon(z,t)=\sum\limits^{\infty}_{n=1}\epsilon_{n}(t)\sin(\pi
nz/L).
\end{eqnarray}
Besides, $\epsilon_{n}(t)$ can be presented as the sum of periodic
and random components
\begin{eqnarray}\label{ch7}
\epsilon_{n}(t)=\epsilon_{n}(t)_{p}+\epsilon_{n}(t)_{r}.
\end{eqnarray}
Substitution of Eqs.(\ref{ch5}) and (\ref{ch7}) into
Eq.(\ref{eq35}) gives the following system of equations:
\begin{eqnarray}\label{ch9}
\frac{\partial ^{2}\epsilon_{n}(t)_{r} }{\partial t^{2}}+
\frac{B}{m}\frac{\partial\epsilon_{n}(t)_{r}}{\partial t}+
\Omega^{+2}_{n}\epsilon_{n}(t)_{r}-
\frac{4b\eta(t)}{\pi n m}=0,\nonumber\\
\frac{\partial ^{2}\epsilon_{n}(t)_{p} }{\partial t^{2}}+
\frac{B}{m}\frac{\partial\epsilon_{n}(t)_{p}}{\partial t}+
\Omega^{+2}_{n}\epsilon_{n}(t)_{p}-\frac{4b\sigma_{0}e^{i\omega
t}}{\pi n m}=0.
\end{eqnarray}
The solution to the second equation is written as
\begin{eqnarray}\label{eq39}
\epsilon_{n}(t)_p=C_{n}e^{i\omega t},
\end{eqnarray}
with $$C_{n}=\frac{b\sigma_{n}}{m}\frac{1}{\left
(i\frac{B\omega}{m}+\Omega^{+2}_{n}-\omega^{2}\right)}. $$ The
solution to the first equation in Eq.(\ref{ch9}) is written in the
form
\begin{eqnarray}\label{ch10}
\epsilon_{n}(t)_{r}=
\frac{1}{m\sqrt{\Omega^{+2}_{n}-\frac{B^{2}}{4m^{2}}}}
\int\limits_{0}^{t}\frac{4b\eta(\tau)}{\pi n}
e^{-B(t-\tau)/2m}\sin\sqrt{\Omega^{+2}_{n}-\frac{B^{2}}{4m^{2}}}(t-\tau)d\tau.
\end{eqnarray}
As is seen, the displacement $\epsilon_{n}(z,t)$ depends on the
random function $\eta(t)$. As a result, the dipole dynamics is
governed by a random function having its own probabilistic
characteristics. This is an essential difference from the standard
string model where the string is subjected only to the influence
of the periodic force.

\section{Internal friction}

Let us calculate the total decrement $\Delta_{t}$ due to the
motion of optically vibrating edge dislocation dipoles by using
the formulated model. Since $\epsilon_{n}(t)$ is a random
function, one has to perform an additional averaging over the
possible realizations of random stresses $\eta(t)$. The mean
energy loss to friction due to optically vibrating edge
dislocation dipole per unit time reads
\begin{eqnarray}\label{ch11}
<\overline{P}>=2\int\limits_{0}^{L}\lim\limits_{T\rightarrow
\infty}\frac{1}{T}<\int\limits_{0}^{T}
(F^{+}+b\eta(t))\frac{{\partial\epsilon(z,t)}}{\partial t}dt>dz,
\end{eqnarray}
where  $<>$ means averaging over random force realizations
ensemble. Generally,
\begin{eqnarray}\label{ch13}
\Delta_{t}=\frac{<\Delta W>}{2<W>}=\frac{N<\Delta W_{d}>}{2<W>} =
\frac{N<\overline{P}>}{2<W>}\frac{2\pi}{\omega},
\end{eqnarray}
where $<W>$ is the averaged total vibration energy per unit
volume,
 N is the total number of
dislocation dipoles per unit volume, $<\Delta W_{d}>$ is the
contribution from the single dislocation dipole. After
substitution of Eq.(\ref{ch7}) into Eq.(\ref{ch11}) the decrement
is written as
\begin{equation}\label{ch15}
\Delta_{t}=\Delta_{r}+\Delta_{p},
\end{equation}
where
$$
\Delta_{r}=\lim\limits_{T\rightarrow\infty}\sum\limits_{l=0}^{\infty}
\frac{16J}{m\omega\pi(2l+1)^{2}}\int\limits_{0}^{T}<\eta(T-S)\eta(T)>
e^{-BS/2m}\left(\cos(\Omega_{l}S)-\frac{B}{2m}\frac{\sin(\Omega_{l}S)}{\Omega_{l}}\right)dS,
$$
$$
\Delta_{p}=\sum\limits_{p=0}^{\infty}\frac{8JB\sigma^{2}_{0}
\omega}{\pi E_{p}(2p+1)^{2}},\quad
E_{p}=\left(D-m\omega^{2}+\frac{T_0
\pi^{2}}{L^{2}}\left(2p+1\right)^{2}\right)^{2}+B^{2}\omega^{2},\quad
J=\frac{b^{2}LN}{<W>}=\frac{b^{2}\Lambda}{2<W>}.
$$
Eq.(\ref{ch15}) describes the contribution to the internal
friction due to dislocation dipoles under the action of both
periodic and random forces. As is known, the decrement describes
how quickly the amplitude of a wave decays. In turn, the
scattering rate, which is necessary to describe the thermal
conductivity, defines the rate at which the incident wave loses
its energy to the dislocation dipoles. It is given by
\begin{equation}\label{eq47}
\tau^{-1}_{d}=\frac{\Delta_{t}\omega}{\pi}.
\end{equation}
Notice that both the decrement and the scattering rate depend on
two sets of parameters. The first one (T$_0$, B, m, D)
characterizes the string  and can be calculated from Eqs.
(\ref{eq3}) and (\ref{eq21}). The second set comes from the
characteristics of the dipole's ensemble such as the density of
dipoles $\Lambda$ and the correlation function
$<\eta(T-S)\eta(T)>$. In general, the dislocation density depends
on the plastic deformation in a sample. Typically, a
phenomenological relation between the plastic strain $\epsilon$
and the dislocation density is used. Experimentally the
plastically deformed samples of Al, Ta, and Nb were studied.
\cite{Pohl:1999,Sahl:2001,Sahl2:2001,Sahl3:2002,Was:2002} The
densities of dislocations in these materials depend on the
specimen preparation and vary in the range $10^{13}-4\times
10^{14}$
m$^{-2}$.\cite{Pohl:1999,Sahl:2001,Sahl2:2001,Sahl3:2002,Was:2002}
In our calculations we try to choose such dipole densities which
give the best fit.

As was mentioned in the introduction, we assume that the
stochastic stress component appears naturally in a high-density
dislocation arrangement. Let us discuss this point in more detail.
Evidently, it is impossible to describe the exact dynamical
behavior of large dislocation ensemble in the complex potential
relief. Instead, one can use the statistical consideration by
introducing the statistical fluctuations of stresses. Notice that
even idealized numerical simulations of the overdamped dynamics of
an array of parallel edge dislocations with long-range
interactions in Ref. \onlinecite{Groma:1998} show an appearance in
this system of the stochastic component of the internal stress
field. According to Ref. \onlinecite{Groma:1998}, for a period of
time shorter than the relaxation time of the dislocation system,
the internal stresses caused by dislocations in the arrangement
can be viewed as a sum of slowly varying mean stress originating
from the internal stress of the smoothed out dislocation
distribution and a rapidly varying, highly irregular function with
a zero mean value that satisfies the requirement of no correlation
at different times or places and that represents the influence of
the nearest neighbors. In this case, the stochastic stress
component has a white noise character.

\begin{figure}[htb]
\includegraphics[width=3.3in]{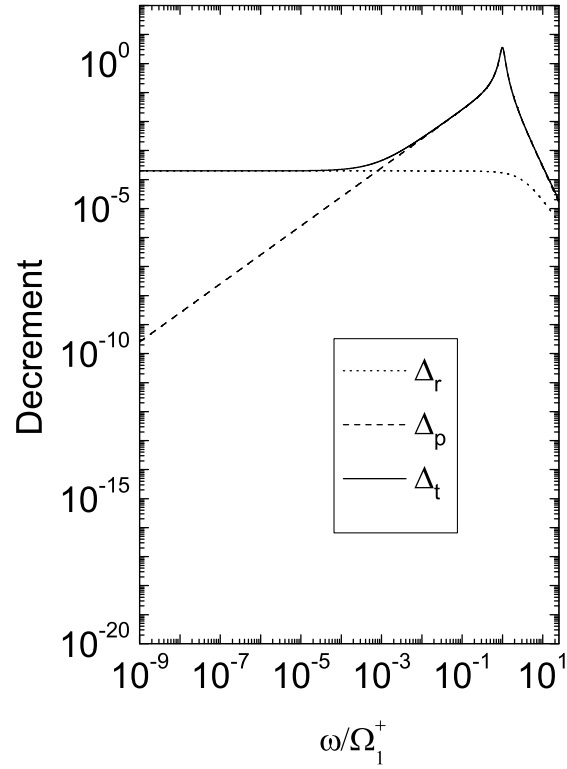}
\vspace{3mm} \caption{The calculated internal friction
 using the parameter set for the 10\% deformed high purity 5N polycrystalline aluminium sample
(Tables I and II) and $\Omega^{+}_{1}=10^{11}$ Hz.} \label{fig4}
\end{figure}

Our calculations show that the white noise does not give a
contribution to the internal friction. As is well known, however,
the white noise concept does not reflect the real physical
processes in complex systems. A possible way to generalize the
consideration is to introduce the colored noise which can be
modeled by exponentially decreasing correlator in the simplest
case. As possible sources of colored noise in dislocation ensemble
one can mention an external force, a long-range interaction
between dislocations, the lattice itself, etc. Actually, we
consider the dichotomic noise which is characterized by
\begin{eqnarray}\label{ch19}
<\eta(t)>=0,\quad
<\eta(t_{1})\eta(t_{2})>=\eta_{0}^{2}e^{-2\alpha\mid
t_{1}-t_{2}\mid},
\end{eqnarray}
where $\alpha$ is a parameter of the corresponding Poisson
process, $\eta_{0}^{2}$ the dispersion.
 Taking into account Eqs.(\ref{ch15}) and (\ref{ch19}), as well as the
fact that for dichotomic noise
$<W>=(\eta^{2}_{0}+\sigma^{2}_{0}/2)/2\mu$, the total decrement
reads
\begin{equation}\label{ch21}
\Delta_{t}=\frac{4\pi J\eta^{2}_{0}}{(4\alpha^{2}m+2\alpha B+D)}
\left[1-\tanh(Y_{t})/Y_{t}\right]\frac{\alpha}{\omega}+\sum\limits_{p=0}^{\infty}\frac{8J
B\sigma^{2}_{0} \omega}{\pi E_{p}(2p+1)^{2}},
\end{equation}
with
$$
Y_{t}=\frac{\pi}{2}\sqrt{\frac{(4\alpha^{2}m+2\alpha
B+D)}{T_0(\pi/L)^{2}}}.
$$
Fig.\ref{fig4} shows the decrement given by Eq.(\ref{ch21}) as a
function of the reduced frequency for parameters of
polycrystalline Al. The corresponding material constants for Al,
Ta and Nb are presented in Table \ref{tab2}. As is seen from
Fig.\ref{fig4}, at low frequencies (up to 10$^{8}$ Hz) the total
decrement is a constant. It is important to note that such
behavior is entirely determined by the contribution from the
random part of stresses. The situation changes drastically near
the resonance frequency where the decrement has a sharp peak and
mainly depends on losses due to the periodic external stresses.
Namely, $\Delta_{p}$ is found to be $10^{4}$ larger than
$\Delta_{r}$ near the resonance frequency. Notice that a similar
behavior is also found for both Ta and Nb. In experiments,
\cite{Pohl:1999,Sahl3:2002,Was:2002} the low-frequency (90kHz
\cite{Pohl:1999,Sahl3:2002} and 1kHz \cite{Was:2002}) internal
friction was studied. Therefore, the region of the constant
decrement is of our main interest here. In addition, in our model
the following condition is fulfilled
\begin{eqnarray}\label{ch23}
D\gg T_0\pi^{2}/L^{2}.
\end{eqnarray}
This is a variant of the so-called "rigid" relief approximation.
\cite{Kamaeva:2004} Moreover, a reasonable restriction on the
dispersion of colored noise can be imposed in the form
$\eta^{2}_{0}/\sigma^{2}_{0}\ll 1$. In view of Eq.(\ref{ch23}),
performing the summation in Eq.(\ref{ch21}) one finally obtains
\begin{eqnarray}\label{ch25}
\Delta_{t}=\frac{8\pi\mu
b^{2}\Lambda}{D}\frac{\eta^{2}_{0}}{\sigma^{2}_{0}}\frac{\alpha}{\omega}+
\frac{2\pi\mu\Lambda Bb^{2}\omega}{D^{2}}.
\end{eqnarray}
The decrement given by Eq.(\ref{ch25}) for different specimens is
presented in Table \ref{tab1}. As is seen, there is a good
agreement with the experimental data
\cite{Pohl:1999,Sahl3:2002,Was:2002} at appropriate choice of the
model parameters. Notice that only two arbitrary model parameters
($\alpha/\omega$ and $\eta^{2}_{0}/\sigma^{2}_{0}$) were used in
our calculations.

It was observed in experiments
\cite{Pohl:1999,Sahl3:2002,Was:2002} that the low-frequency
internal friction does not depend on either the temperature or the
frequency at low temperatures. In the framework of our model this
behavior can be explained only when both the interaction between
dislocations (in a dipole) and the colored noise are present.
Namely, the colored noise gives a dominant contribution (in
comparison with the Granato-Lucke mechanism) to the decrement at
low frequencies. At the same time, the athermal long-distance
interaction (determined completely by the structure of the
dislocation ensemble) provides the athermal behavior of the
low-frequency internal friction at low temperatures. Indeed, this
interaction ensures the validity of the "rigid" relief
approximation (see Eq.(28)) when both the temperature-dependent
terms (containing the damping parameter $B$) and the terms
containing the tension and the mass of the string are negligible
in Eq.(27). Let us note that a similar role of the long-range
interactions was earlier discussed in Ref.
\onlinecite{Kustov:1999} in a study of solid solutions. In
particular, it was shown that the long-range interaction of
dislocations with solute atoms distributed in the bulk of the
crystal supports the athermal behavior of the nonlinear
dislocation strain-amplitude-dependent internal friction in solid
solutions. \cite{Kustov:1999,Gremaud:2004}

We also assume the linear response of
the dislocation array to the external periodic forces which means
$\alpha/\omega$=const, $\eta^{2}_{0}/\sigma^{2}_{0}$=const.
Without loss of generality one can fix the first constant as unity
at numerical evaluations.
Notice that, strongly speaking, these assumptions do not provide
the independence of the decrement from frequency.
Actually, this is true only for low-frequency asymptotic of
the contribution to the decrement from random forces.
Indeed, as is seen from Fig.1 the decrement $\Delta_r$ decreases
rapidly at high frequencies.

\section{Thermal conductivity}

\begin{figure}[htb]
 \includegraphics[width=3.3in]{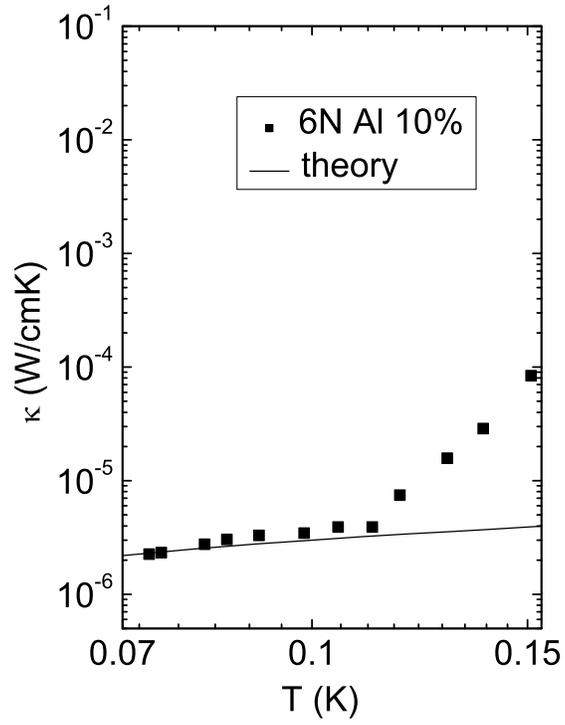}
\vspace{3mm} \caption{\label{fig1} Thermal conductivity for Al.
Experimental data denoted by squares are from Ref. 3. The solid
line represents the theoretical curve for $\Lambda_{Al}=6\times
10^{12}$ m$^{-2}$ using the parameter set from Table I. The
resonance frequency for Al is $\Omega_{1}^{+}=10^{11}$ Hz. }
\end{figure}

\begin{figure}[htb]
\includegraphics[width=3.3in]{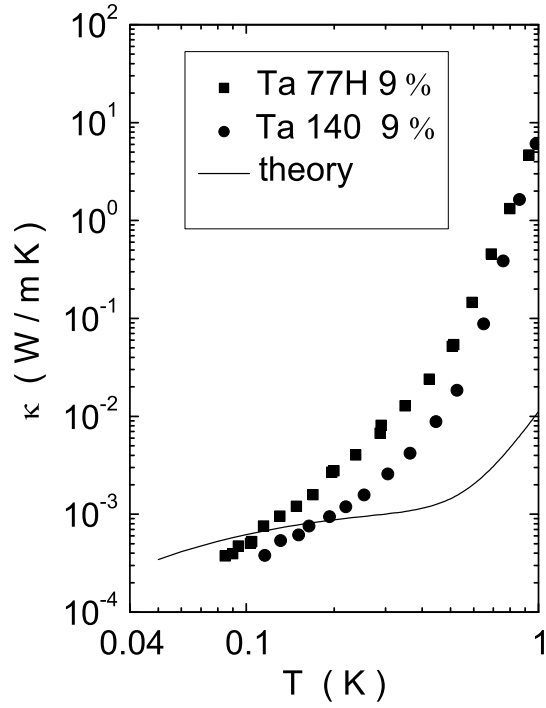}
\vspace{3mm} \caption{Thermal conductivity for different Ta
samples. Experimental data denoted by squares and circles are from
Ref. 2. The solid line represents the theoretical curve for
$\Lambda_{Ta}=2\times 10^{14}$ m$^{-2}$ using the parameter set
from Table I. The resonance frequency for Ta is
 $\Omega_{1}^{+}=2\times 10^{11}$ Hz.} \label{fig2}
\end{figure}

\begin{figure}[htb]
\includegraphics[width=3.3in]{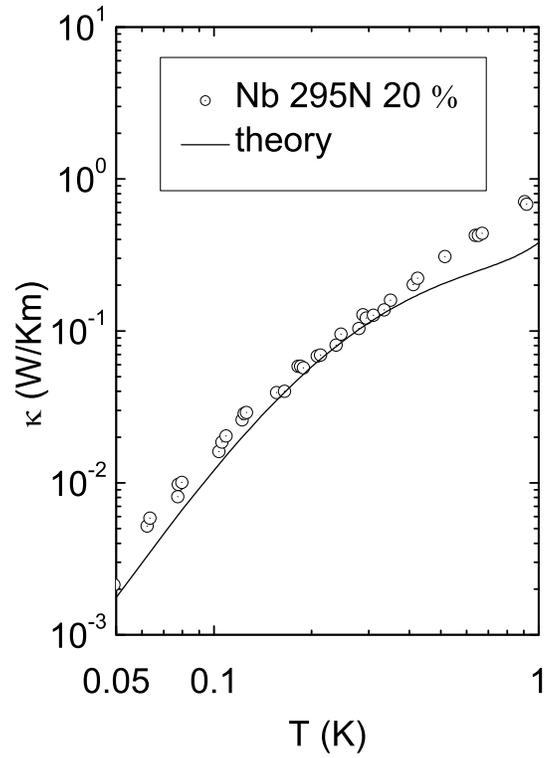}
\vspace{3mm} \caption{Thermal conductivity for Nb. Experimental
data denoted by circles are from Ref. 4. Solid line represents the
best fit with $\Lambda_{Nb}=10^{13}$ m$^{-2}$ using the parameter
set from Table I. The resonance frequency for Nb is
$\Omega_{1}^{+}=5\times10^{11}$ Hz.} \label{fig3}
\end{figure}

As is well-known, at low temperatures the main contributions to
the thermal conductivity come from the lattice defects and
boundary scattering. Within the framework of the relaxation time
description and the Debye approximation the lattice thermal
conductivity is given by
\begin{eqnarray}\label{eq49}
\kappa =
\sum\limits_{j=1}^{3}(1/6\pi^{2}v_{j})\int\limits_{0}^{\omega_{Dj}}
\tau_{j}(\overline{\omega}_{j})\overline{\omega}^{2}_{j}C_{j}
(\overline{\omega}_{j})d\overline{\omega}_{j},
\end{eqnarray}
where the sum includes the longitudinal and two transverse
acoustical phonon modes. $\omega_{Dj}$ is the Debye frequency,
$v_{j}$ the sound velocity, $\tau_{j}$ the total phonon relaxation
time and $C_{j}$ the specific heat of the phonon mode $j$.
Assuming that all phonon modes are scattered almost equally by the
dislocation dipoles, the total relaxation rate $\tau^{-1}$ is
written as
\begin{eqnarray}\label{eq51}
\tau^{-1}=\tau^{-1}_{c}+\tau^{-1}_{d},
\end{eqnarray}
where $\tau^{-1}_{c}$ is the relaxation rate of the boundary
scattering, and $\tau^{-1}_{d}$ is the relaxation rate due to the
dynamic phonon-dipole interaction. As another approximation, we
will use the averaged sound velocity $\overline{v}$ instead of
sound velocity of the phonon mode $j$. In other words, we drop the
polarization effects. It is also convenient to use the
dimensionless variable $x=\hbar\overline{\omega}/kT$ so that
Eq.(\ref{eq49}) takes the form
\begin{eqnarray}\label{eq53}
\kappa =
(k^{3}/2\pi^{2}\overline{v}\hbar^{3})T^{3}\int\limits_{0}^{\theta/T}
x^{2}\tau(x)C(x)dx,
\end{eqnarray}
where $\theta$ is the Debye temperature. As is well-known, the
boundary scattering gives a constant $\tau^{-1}_{c}$ which is
determined by geometry of samples. This contribution can be
estimated by using the known Casimir formula. $\tau^{-1}_{d}$ is
the function of both the decrement $\Delta_{t}$ and the frequency
$\omega$ in accordance with Eq.(\ref{eq47}). The results of
numerical calculations for all the materials are shown in
Figs.2-4. The dislocation length is chosen to be $10^{-6}$ m in
agreement with the acoustic experiments while the dipole
separation $l$ varies from one material to another to get the
necessary resonance frequency. The best fits were found for the
densities $\Lambda_{Al}=6\times 10^{12}$ m$^{-2}$ and
$\Lambda_{Nb}=10^{13}$ m$^{-2}$. The calculated curve for Ta is
shown in Fig.4 with $\Lambda_{Ta}=2\times 10^{14}$ m$^{-2}$.
Notice that these densities of dipoles differ from those used in
the description of the internal friction experiments because the
thermal conductivity and internal friction were actually measured
on different
samples.\cite{Pohl:1999,Sahl:2001,Sahl2:2001,Sahl3:2002,Was:2002}
Therefore, to clarify the problem it would be desirable to measure
both the internal friction and the thermal conductivity on the
same samples under the same conditions (to avoid stress relaxation
and preserve the initial characteristics of dislocation
ensembles). It is important to note that the influence of
dichotomic noise on the thermal conductivity is found to be
negligible. The reason is that the main contribution to the
thermal conductivity comes from the region near the resonance
frequency of optically vibrating dislocation dipoles. In this
case, the term due to the colored noise can be disregarded in the
scattering rate ( in Eq.(\ref{eq47}) $\Delta_{r}$ becomes much
smaller than $\Delta_{p}$, see Fig.1). Thus, in contrast to the
internal friction, the dichotomic noise does not influence the
thermal conductivity. As is seen from Figs.\ref{fig1}-\ref{fig3},
there is a good agreement with the experimental data at low
temperatures. Notice that a rapid increase of experimental curves
with temperature begins in the region where the electronic
contribution becomes essential due to an increase of a number of
normal electrons in superconducting samples.

\section{Summary}

In this paper, we have shown that the presence of a stationary
random process in dislocation ensemble caused by the external
perturbation markedly modifies the decrement in the low-frequency
range. First, the decrement does not depend on either frequency or
temperature at low temperatures and, second, its value becomes
much higher in comparison with the Granato-L\"ucke classical case
of periodically affected dislocations. Both these findings agree
well with the experimental data. At the same time, an explicit
type and probabilistic characteristics of the proposed stochastic
process are still not understood because dislocation ensembles
show a very complicated dynamics. In fact, by introducing the
dichotomic noise we simplify the problem of the nonlinear dynamics
of dislocation dipoles to the linear one. We have considered the
influence of both external force and colored noise on the internal
friction. We have also calculated the contribution to the thermal
conductivity due to dislocation dipoles. The concept of
dislocation dipoles allows us to obtain the required resonance
frequency and, as a consequence, to describe the experimental
results for all three plastically deformed metals. It was found
that the dichotomic noise does not play any essential role in the
thermal conductivity.

Nevertheless, it is still impossible to say that the colored noise
in the dislocation ensemble is the only way to explain the
experimentally observed damping in Refs.
\onlinecite{Pohl:1999,Sahl3:2002}, and \onlinecite{Was:2002}. The
purpose of our paper was to draw attention to the fact that this
mechanism gives an important contribution to the damping at low
temperatures. A good agreement with the experiment was obtained by
appropriate choice of two arbitrary parameters noticed above.
There are other known mechanisms of damping which could be of
importance to explain the experimental data.

One of them was suggested in Ref. \onlinecite{Pohl:1999} where the
tunneling of some entities (presumably dislocations or dislocation
kinks) was discussed as a possible reason for a large internal
friction observed in plastically deformed metals at low
temperatures. Theoretically, the structures, energy barriers,
effective masses, and quantum tunneling rates for dislocation
kinks and jogs in copper screw dislocations were determined in
Ref. \onlinecite{Vegge:2001} by using the molecular-dynamics
calculations. It was found that for dislocation kinks in screw
dislocations both the energy barrier and the effective mass are
markedly reduced so that tunneling should occur readily. The kink
tunneling is very sensitive to both the effective mass and the
Peierls-like barrier for migration along the dislocation which, in
turn, depend on the spatial extent of the kink. In particular, for
wide kinks the barrier grows exponentially with decreasing kink
width \cite{Vegge:2001}. Hence, even a minor change of the kink
width results in the substantial variation of the WKB factor
(suppressing quantum tunneling). Notice that the value of the kink
width used in Refs. \onlinecite{Sahl:2001,Sahl2:2001,Sahl3:2002}
for Al, Ta and Nb ($w=5b\sim 1.4$ nm) is about three times shorter
than that in Ref. \onlinecite{Vegge:2001} for Cu. Besides,
typically kink tunneling is suppressed by dissipation (see, e.g.,
Ref. \onlinecite{Hikata:1985}). In addition, unfortunately both
the resonance frequency and the scattering cross-section of the
tunneling process are not yet available. This gives no way of
calculating the contribution to the thermal conductivity within
the kink tunneling model. Therefore, additional investigations
within the tunneling kink model are necessary to describe a series
of experiments in plastically deformed Al, Ta and
Nb.\cite{Pohl:1999,Sahl:2001,Sahl2:2001,Sahl3:2002,Was:2002}

As another possible mechanism, the concept of kinks moving in the
secondary Peierls potential was suggested for the description of
observed experimental data.\cite{Sahl:2001,Sahl2:2001,Sahl3:2002}
Despite a good agreement between the results of the kink model and
the experiment, two comments are in order. First, it was found
that a good fit for the thermal conductivity takes place only for
a surprisingly large number of kinks ($\sim 500$) per dislocation
line of length $~10^{-6}$m. In this case, the separation between
kinks turns out to be 2nm. This value is comparable with the kink
width $w\sim 1.4$ nm used in Refs.
\onlinecite{Sahl:2001,Sahl2:2001,Sahl3:2002}. At the same time,
the radiation mechanism proposed in Ref. \onlinecite{Hikata:1974}
is essentially based on the long-range interaction between kinks
and is only valid when the separation between kinks is large
compared to the kink width (otherwise the interaction between
kinks may change drastically) \cite{Seeger:1966}. Second, within
the kink model the internal friction significantly (as a third
power) depends on the Peierls stress whose reliable estimation is
a complex experimental task \cite{Friedel:1964}. As a final
remark, at large densities the interaction between dislocations is
of importance, hence the complex dislocation dynamics is to be
expected.

\begin{acknowledgments}
This work has been supported by the Heisenberg-Landau programme.
\end{acknowledgments}

\begin{table}[htb]
  \centering
 \caption{\label{tab2} Materials constant for Al (from Ref. 3), Ta (from Ref. 2), and Nb (from Ref. 4).}
 \includegraphics[width=4in]{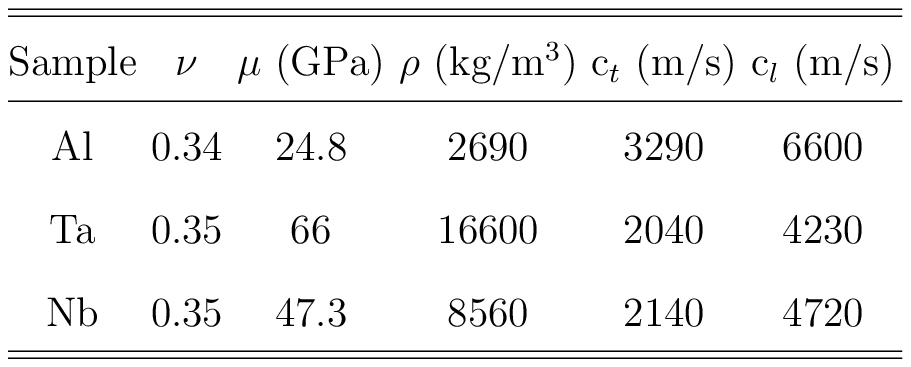}
\end{table}

\begin{table}[htb]
 \centering
\caption{\label{tab1} Low frequency internal friction calculated
according to Eq.(29) for different metals.} \epsfxsize=12cm
\includegraphics[width=4in]{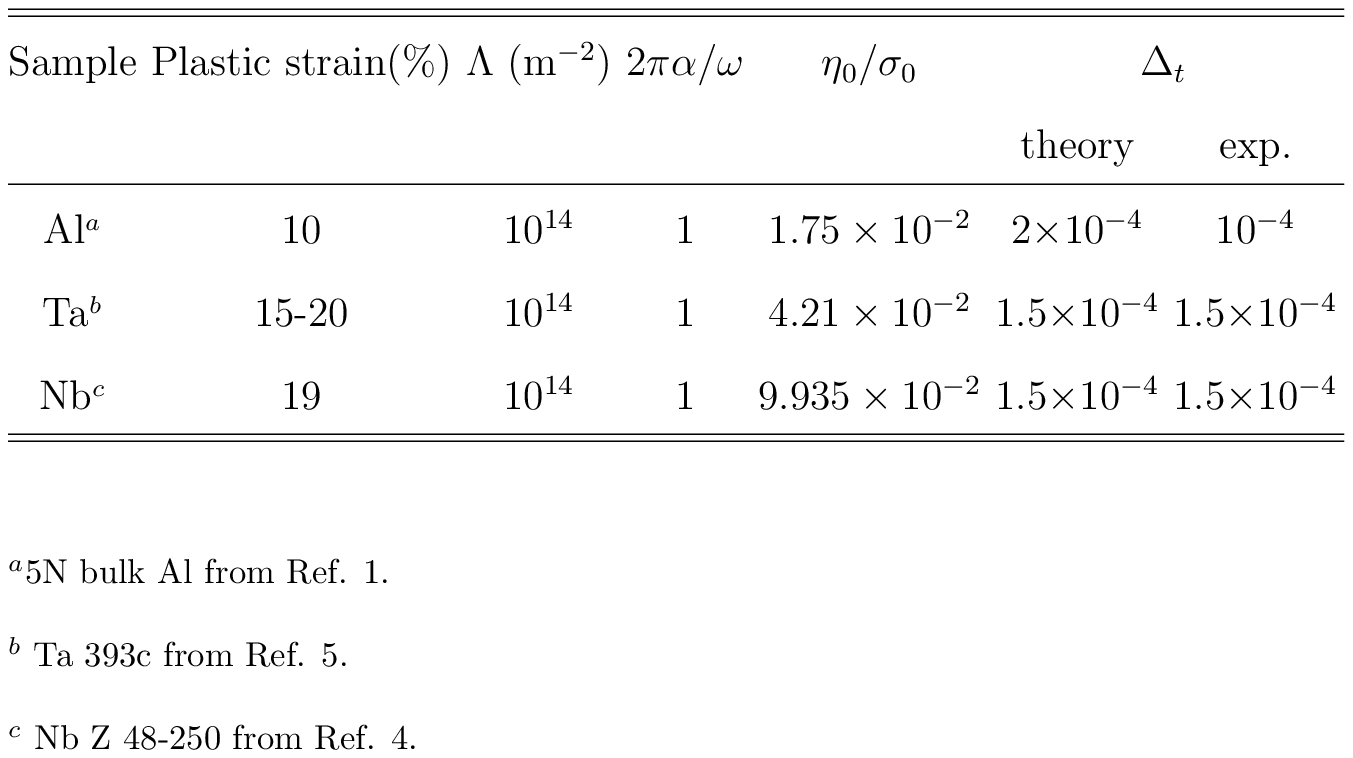}
\end{table}

\end{document}